# Optimizing Convolutional Neural Networks for Chronic Obstructive Pulmonary Disease Detection in Clinical Computed Tomography Imaging

Tina Dorosti[a,b,c,*], Manuel Schultheiss[a,b,c], Felix Hofmann[c], Johannes Thalhammer[a,b,c,d], Luisa Kirchner[c], Theresa Urban[a,b,c], Franz Pfeiffer[a,b,c,d], Florian Schaff[a,b], Tobias Lasser[b,e] and Daniela Pfeiffer[c,d]

[a]Chair of Biomedical Physics, Department of Physics, School of Natural Sciences, Technical University of Munich, Garching, 85748, Bavaria, Germany

[b]Munich Institute of Biomedical Engineering, Technical University of Munich, Garching, 85748, Bavaria, Germany

[c]Department of Diagnostic and Interventional Radiology, School of Medicine and Health, Klinikum rechts der Isar, Technical University of Munich, Munich, 81675, Bavaria, Germany

[d]Institute for Advanced Study, Technical University of Munich, Garching, 85748, Bavaria, Germany

[e]Computational Imaging and Inverse Problems, Department of Computer Science, School of Computation, Information, and Technology, Technical University of Munich, Garching, 85748, Bavaria, Germany

## ARTICLE INFO

*Acknowledgements*:
Funded by the Federal Ministry of Education and Research (BMBF) and the Free State of Bavaria under the Excellence Strategy of the Federal Government and the States, the German Research Foundation (GRK2274), as well as by the Technical University of Munich–Institute for Advanced Study.

The authors report no conflicts of interest.

## ABSTRACT

We aim to optimize the binary detection of Chronic Obstructive Pulmonary Disease (COPD) based on emphysema presence in the lung with convolutional neural networks (CNN) by exploring manually adjusted versus automated window-setting optimization (WSO) on computed tomography (CT) images. 7,194 contrast-enhanced CT images (3,597 with COPD; 3,597 healthy controls) from 78 subjects were selected retrospectively (01.2018-12.2021) and preprocessed. For each image, intensity values were manually clipped to the emphysema window setting and a baseline 'full-range' window setting. Class-balanced train, validation, and test sets contained 3,392, 1,114, and 2,688 images. The network backbone was optimized by comparing various CNN architectures. Furthermore, automated WSO was implemented by adding a customized layer to the model. Repeated inference (n=7) on the test set showed that the DenseNet was the most efficient backbone and achieved a mean AUC of 0.80 [0.76, 0.85] without WSO. Comparably, with input images manually adjusted to the emphysema window, the DenseNet model predicted COPD with a mean AUC of 0.86 [0.82, 0.89]. By adding a customized WSO layer to the DenseNet, an optimal window in the proximity of the emphysema window setting was learned automatically, and a mean AUC of 0.82 [0.78, 0.86] was achieved. Detection of COPD with DenseNet models was improved by WSO of CT data to the emphysema window setting range.

*Corresponding author
tina.dorosti@tum.de (T. Dorosti)
ORCID(s): 0000-0002-5747-3250 (T. Dorosti)

## 1. Introduction

Chronic Obstructive Pulmonary Disease (COPD) is a group of respiratory diseases impairing the lung structure, such as emphysema [1]. With 3.23 million deaths recorded globally in 2019, COPD is among the leading causes of death worldwide [2]. In addition to increased mortality rates directly correlating with the disease, patients with COPD preconditions are at a higher risk for all-cause mortality [3][4][5]. With early detection and intervention, COPD's prevalence and negative impacts can be decreased [3].

Spirometry is a readily available pulmonary function test for COPD detection and is often utilized to categorize the disease progression into four stages (GOLD I-IV) [1]. Although spirometry reliably detects advanced stages of COPD, false negative results dominate in the early stages [1][6]. Spirometry tests are highly technique-dependent and susceptible to minor mistakes in following breathing instructions [7]. Moreover, patients diagnosed with COPD at the same GOLD stage have shown drastic morphological differences in the lung structure [8].

Alternatively, COPD is detected with an X-ray Computed Tomography (CT) scan, where detailed three dimensional morphological information about the lung structure is obtained in Hounsfield Unit (HU) values. This information on phenotypic abnormalities and patterns of morphological changes reflecting emphysema allows for detecting and controlling disease progression even in the early stages. In 2015, the Fleischner Society introduced a disease progression scale based on the pattern of abnormalities in CT data corresponding to COPD and emphysema subtypes [8].

In recent years, increasing large-scale COPD studies [9][10] in parallel with advancing machine learning models have made Convolutional Neural Networks (CNNs) a popular tool for detection of COPD [11][12][13][14][15][16]. Radiomics approach for COPD detection with deep learning models has also shown promising results by extracting CT-based features and accounting for this information in the deep learning model [17][18][19]. Although such high-level features improve the model outcome, their complexity adds to the 'black-box' nature of machine learning algorithms. Deep-learning models need to be interpretable and explainable before practitioners can accept and implement these models in the healthcare system [16][20]. Therefore, despite their success, these models are still not ready for integration into a computer-aided clinical workflow for efficient COPD diagnosis.

The impact of image preprocessing, such as geometry-based transforms from computer vision [21] and conventional preprocessing methods [22] have been explored for COPD detection to improve CNN and radiomic-based models, respectively. CNN research in other medical imaging tasks has demonstrated the benefits of incorporating clinically relevant steps in the model workflow. In particular, optimizing window-setting parameters of input CT images can improve the results [23][24][25]. However, to the best of our knowledge, implementing automated preprocessing steps to adapt the clinical workflow process of window-setting optimization has not been explicitly explored in the extant literature on COPD detection with CNNs.

We hypothesize that COPD detection with CNN models can be improved by adapting preprocessing based on existing radiological knowledge, specifically through window-setting optimization (WSO). In this exploratory study, we aim to optimize the binary detection of COPD based on emphysema presence in the lung with CNNs by exploring the effects of manually adjusted versus automated WSO on CT images.

## 2. Methods

### 2.1. Dataset

7,194 contrast-enhanced CT (CECT) images from 78 subjects and 5,086 non-contrast CT (NCCT) images corresponding to 21 subjects were selected retrospectively (01.2018-12.2021), as shown in Figure 1. CT images acquired both with and without contrast material were considered: CECTs contribute to numerous COPD diagnoses from incidental findings, and NCCTs are commonly used to evaluate and control disease progression in patients with COPD [26]. All procedures were performed in compliance with the relevant laws of our institutional ethics review board. Approval from the ethics committee (IRB code: 87/18 S received on 03.2018) and patient informed consent were obtained. The CT scans were first anonymized and then graded based on the Fleischner Score categories of centrilobular emphysema by three expert radiologists with 4-12 years of experience for the CECT data, and by an in-training radiologist (LK) for the NCCT data [8]. Contrary to the spirometry-based GOLD staging, the Fleischer system is defined based on the morphological characteristics of emphysema visualized in CT data. Therefore, Fleischner scores were chosen as the ground truth labels because they provide a more accurate description of the observable disease patterns in the CT images and were deemed more suitable for this computer vision task. Scans with scores greater than 'mild' were considered as the COPD class for the binary classification task. Scans with 'moderate' scores were further annotated on image level by another radiologist (FH) to distinguish images presenting COPD. Emphysematous areas were not segmented, and emphysema-thresholded maps were not generated. CT images from subjects suffering from COPD ($n_{CECT}$=3,597, $n_{NCCT}$=2,543) were considered as the COPD class, and the remaining CT images ($n_{CECT}$=3,597, $n_{NCCT}$=2,543) were assigned to the no COPD class. Datasets were class-balanced and selected on image level such that additional slices from the larger class were randomly removed to match the number of slices in both classes. Train, validation, and test sets included 3,392, 1,114, and 2,688 images from the CECT set, respectively. The NCCT data was reserved to test the robustness of the model on out-of-distribution images acquired without iodinated contrast material. Patient demographics for each set are given in Table 1.

## 2.2. Data preprocessing

Each image was segmented to the lung region, its intensity values clipped to the respective window setting, and normalized. Window settings are given by the window width and level (WW, WL) HU. Manual optimization of the window setting was carried out by reviewing relevant literature on common HU ranges associated with emphysema, and the emphysema window (124, -962) HU [27] was selected for the classification of COPD such that the maximum and minimum intensity values were set to -900 HU and -1024 HU, respectively. Furthermore, a 'full-range' windowing (2048,0) HU was defined based on the minimum, -1024 HU, and the maximum, 1024 HU, intensity values recorded over all images. The full-range window was considered the baseline window setting. Figure 2 shows CECT examples of images from both classes preprocessed to the full-range and the emphysema windows. Additionally, the same example images are shown in the conventional radiological window for viewing lungs (1500, -700) HU [8] as a reference standard. It can be observed that emphysematous patches of low attenuation have more contrast in the images preprocessed to the emphysema window.

## 2.3. Network Architectures

All models were implemented in TensorFlow (2.4.0) [28] and compiled with binary cross entropy loss and the Adam optimizer [29]. Reducing the learning rate by a factor of ten and early stopping were scheduled over 15 and 50 epochs, respectively, if the validation loss did not decrease. 256 by 256 pixel CT slices were used as input data for the models.

### 2.3.1. Backbone Comparison

Taking the reported CNN models with promising results in COPD detection [14][15] as a starting point, DenseNet-121 [30], EfficientNetB2 [31], and ResNet-34 [32] architectures were examined to select the model with the best performance. The number of trainable parameters for each model was respectively, 6.96, 7.70, and 15.7 million parameters. The models were trained and tested on images linearly clipped to the full-range and the emphysema window settings to analyze the influence of the window-setting preprocessing on binary COPD detection. Based on the results presented in Table 2 and section 3.1, DenseNet-121 (plain DenseNet) was chosen as the backbone architecture. Figure 3 details the selected DenseNet architecture.

### 2.3.2. DenseNet$_{WSO}$

A WSO layer was added to the plain DenseNet, as suggested by [24], to create DenseNet$_{WSO}$. Here, only the ReLU activation function is considered for the WSO layer, as it consistently outperformed the sigmoid variant. Therefore, the WSO layer depicted in Figure 3 consisted of a 1x1 convolution layer followed by a ReLU activation. The ReLU acted as a windowing function and was trained to find an optimal window setting for the detection task. The WW and WL values related to the learnable weight (w) and bias (b) parameters of the ReLU function, taken from [24] with correction,

$$f_{ReLU}(x) = max(min(wx + b, U), 0), \quad \text{where}$$
$$w = \frac{U}{WW}, \quad b = \frac{U}{WW}(\frac{WW}{2} - WL). \tag{1}$$

The upper bound for the ReLU windowing function, U=1, was set to achieve learned window settings ranging between zero and one. The DenseNet$_{WSO}$ model was trained to converge to an optimal window setting after initialization to either the full-range or the emphysema window settings while simultaneously adjusting learnable parameters of the DenseNet block for the detection task. Initialization of the WSO layer was carried out by defining the learnable parameters for each window setting respectively. All input images for DenseNet$_{WSO}$ were given in the full-range window and normalized. The optimal window settings learned by the model were calculated with (1).

### 2.3.3. DenseNet$_{FNF}$

Based on the reports in section 3.2, the DenseNet$_{WSO}$ model struggled to converge to optimal results for simultaneous COPD detection and window-setting optimization: To stabilize the learned window setting over all runs, the DenseNet$_{WSO}$ model was first trained with the learnable parameters from the WSO layer frozen

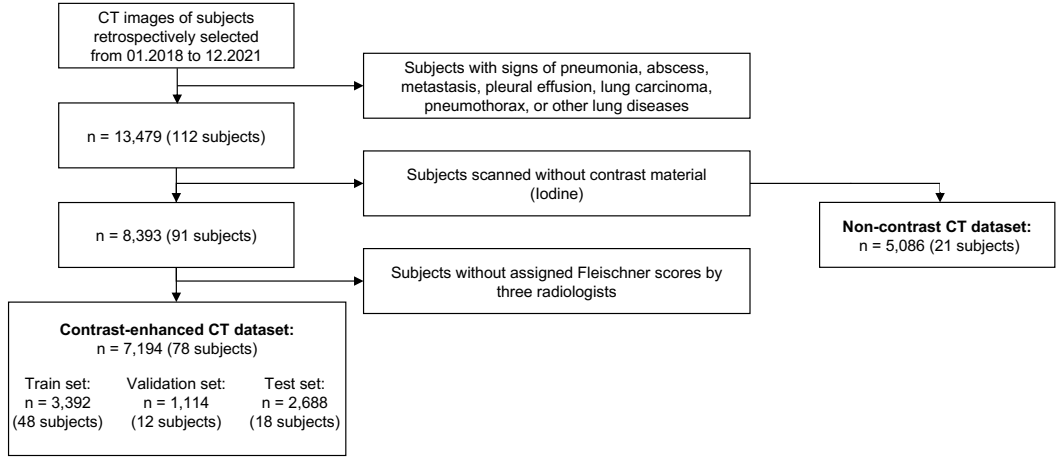

**Figure 1:** Flowchart describing the data selection process. 7,194 contrast-enhanced CT images, and 5,086 non-contrast CT images from our clinic were selected retrospectively and anonymized for the binary detection of Chronic Obstructive Pulmonary Disease (COPD).

**Table 1**
Subject Demographics of the Contrast-enhanced CT (n = 78) and the Non-contract CT (n = 21) Datasets

| Subset | Train | | Validation | | Test | | Non-contract CT Test | |
|---|---|---|---|---|---|---|---|---|
| Parameter | No COPD | COPD | No COPD | COPD | No COPD | COPD | No COPD | COPD |
| Male | 21 | 5 | 6 | 1 | 10 | 5 | 10 | 9 |
| Female | 20 | 2 | 4 | 1 | 3 | 0 | 2 | 0 |
| Age [1] | 62.7 | 70.3 | 63.9 | 71.0 | 64.9 | 75.2 | 72.6 | 65.2 |
| (years) | [34, 91] | [55, 80] | [49, 80] | [70, 72] | [31, 77] | [65, 83] | [54, 87] | [54, 81] |

[1] Note. Age is given in terms of each set's mean and age range [youngest, oldest]

and fixed to the initialized settings. Then, the model was further trained with the unfrozen WSO layer, which allowed its parameters to adjust for the optimal window setting. The same model was also trained continuously for a third round with the learnable parameters of the WSO layer frozen. This training sequence with frozen, not frozen, and frozen (FNF) WSO layer learnable parameters is called DenseNet$_{FNF}$, and attempts to refine model training for the given tasks in multiple stages.

## 2.4. Evaluation Metrics

A hold-out test set was favored over k-fold cross-validation due to the limited number of unique CT-level data points. Subjects were split into train and test sets on a CT level to consider an equal distribution of the diseased subjects with different severity of emphysema for the train and test splits. Reported results on the held-out CECT test set, as well as the out-of-distribution NCCT test set, demonstrated that our choice of a train-validation-test split method did not impede model performance.

Models were initialized randomly and trained from scratch for seven runs; each model run was inferred once with the test data. The Receiver Operating Characteristics (ROC) curve and the area under the ROC curve (AUC) were used to assess the models' performance. Utilizing different threshold choices, the AUC alleviates the ambiguity regarding maximizing sensitivity or (1 - specificity) for smaller sample sizes [33]. The Scikit-learn library (1.2.0) was used to generate the ROC curves, choose optimal thresholds for each curve, and calculate the respective AUC values and 95% confidence Intervals (CI) [34].

## 3. Results

This section presents the binary COPD detection results of the backbone comparison and the three DenseNet variants on the CECT test set's 2,688 images, as well as NCCT test set's 2,543 images. The AUC values for the train and validation process of the CECT images are provided in Table S1 for DenseNet variants. To demonstrate the comparability and the general applicability of our proposed model with the other clinically relevant definition of COPD, the CECT test set was additionally analyzed with the GOLD score ground truth labels, such that GOLD scores above zero were considered as the COPD class ($n_{Healthy} = 1,861$, $n_{COPD} = 827$ with $n_{GOLD-II} = 219$, $n_{GOLD-III} = 608$). These results are presented in Table S2. External public datasets were not considered due to a lack of Fleischner score ground truth labels.

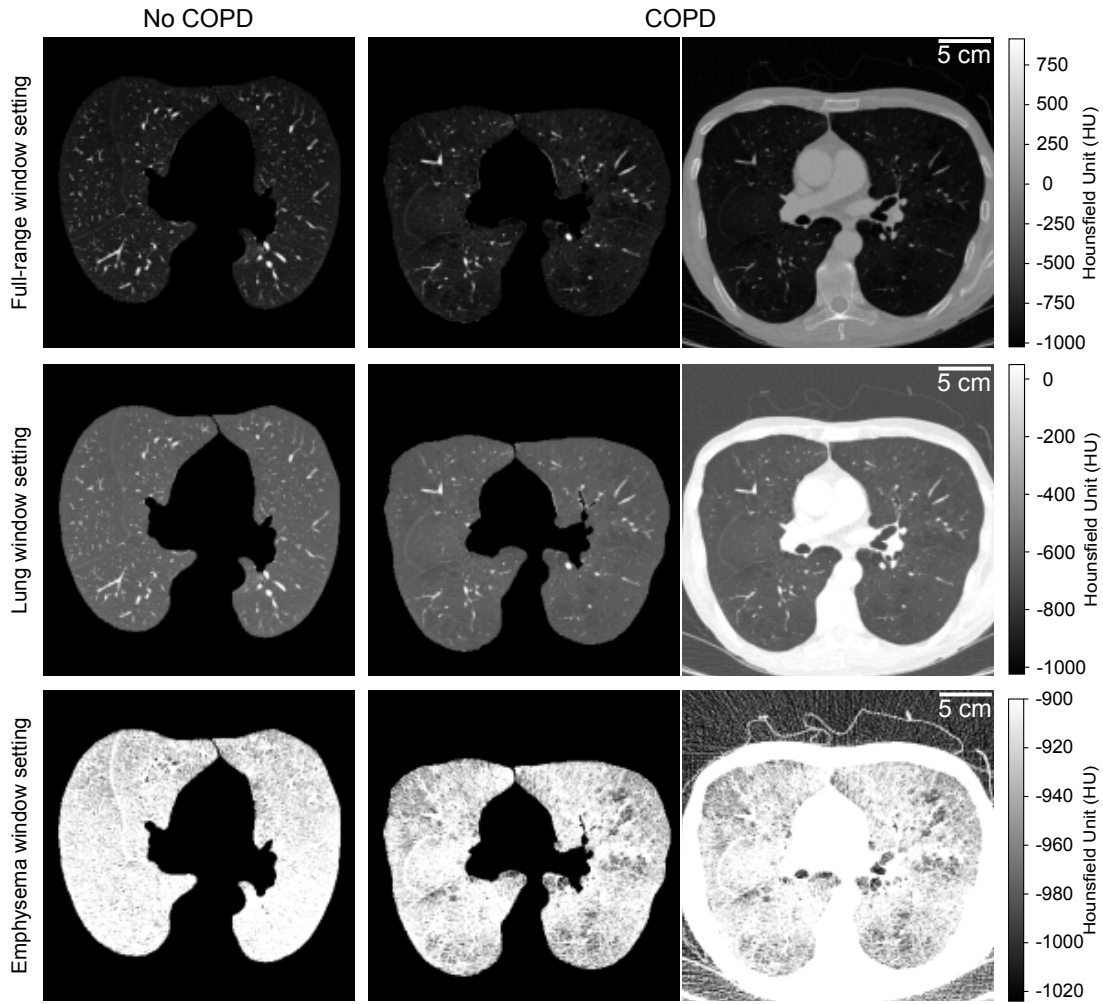

**Figure 2:** Example images from the Chronic Obstructive Pulmonary Disease (COPD) and no COPD class preprocessed to full-range, lung, and emphysema window settings. The images preprocessed to the lung window setting serve as a reference to the common radiological window used to view the lungs. The no COPD image is segmented to the lung region and corresponds to a healthy subject. The COPD image is shown in both segmented and original form and corresponds to a subject with a Fleischner score of 'advanced' centrilobular emphysema. The nonhomogeneous patches of low attenuation corresponding to emphysema are more contrasted in the emphysema-clipped image. All images belong to the contrast-enhanced CT dataset and include a contrast medium (Iodine).

## 3.1. Backbone Architecture and Manual WSO

Table 2 provides the AUC values for the architectural backbone comparison on the CECT and NCCT test sets over seven runs. Preprocessing the input data to the emphysema window consistently improved all model performances for both CECT and NCCT test data. EfficientNetB2 and DenseNet-121 showed higher AUC values compared to ResNet-34 for the CECT data, whereas the ResNet-34 and DenseNet-121 demonstrated more robustness for the out-of-distribution NCCT data. Regarding computational efficiency, ResNet-34 and EfficientNetB2 had an approximate mean training time of 90 minutes per run, whereas DenseNet-121 had a mean training time of roughly 30 minutes per run. Consequently, DenseNet-121 (plain DenseNet) was chosen as the backbone architecture for the task at hand.

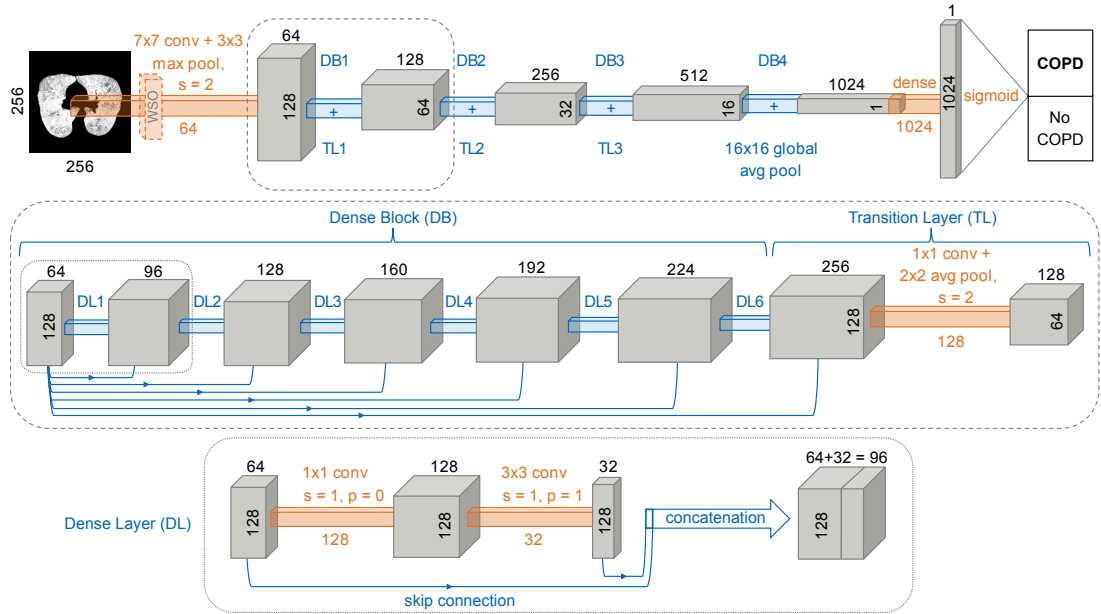

**Figure 3:** DenseNet architecture with 121 layers used for binary detection of Chronic Obstructive Pulmonary Disease (COPD). The model constituents, Dense Block (DB), Dense Layer (DL), and Transition Layer (TL), are expanded in detail. The convolution (conv) and pooling (pool) layers are described by their stride (s) and padding (p) parameters. DenseNet-characteristic skip connections are shown in the DB and DL. The model had a growth rate of 32. The window-setting optimization (WSO) layer consisted of a 1x1 convolution layer followed by a Rectified Linear Unit (ReLU) activation and was used for the automatic optimization of the window settings in the DenseNet$_{WSO}$ and DenseNet$_{FNF}$ implementations. The architecturally specific vertical digits for each box represent the side length dimensions, and the numbers over each block correspond to the number of channels.

Using the plain DenseNet model, image-level ROC plots with corresponding AUC values were compared between full-range and emphysema window settings in Figure 4 for the CECT data and in Figure S1 for the NCCT data. Since the test set had balanced images from both classes of no COPD and COPD, the chance diagonal was used as a visual guide to mark the AUC value of 0.5. Plain DenseNet results in Figure 4 and Figure S1 show that clipping data to the emphysema window setting consistently results in higher values and narrower 95% CI (mean AUC$_{CECT}$ = 0.86 [0.82, 0.89], mean AUC$_{NCCT}$ = 0.82 [0.80, 0.84]) in comparison to the full-range window setting (mean AUC$_{CECT}$ = 0.80 [0.76, 0.85], mean AUC$_{NCCT}$ = 0.78 [0.73, 0.83]). Among all plain DenseNet results, the last model run, with the CECT input images preprocessed to the emphysema window setting, led to the highest AUC value of 0.91.

## 3.2. Automatic WSO

The window-setting values in Table 3 correspond to the mean and 95% CI values for WW and WL over the seven runs of each arrangement. The information in Table 3 is independent of the inference data set, as the learned window-setting values are fixed model-specific parameters after a completed training run. The learned WW and WL parameters were calculated from the weights and bias values of the WSO layer using (1). Figure 5 shows the learned and the corresponding initialization window setting for each WSO model. Note that the window settings used for the initialization of WSO models were the same as the parameters used for preprocessing the inputs to the plain DenseNet.

A shift towards the lower end of the HU range in all learned window settings is noticeable, as given in Table 3 and Figure 5. Over seven runs, the mean learned WL decreased more drastically for models initialized to the full-range window setting. The observed trends suggest a convergence towards the standard emphysema window setting for the learned WW and WL parameters by DenseNet$_{WSO}$ and DenseNet$_{FNF}$ when initialized to the full-range window setting. Between the two models, DenseNet$_{FNF}$ learned a window setting closer to the emphysema window setting regardless

**Table 2**

Mean Area Under the Receiver Operating Characteristics Curve (AUC) for Backbone Architecture Comparison (n=7)

| Test Set | Contrast-enhanced CT | | Non-contrast CT | |
|---|---|---|---|---|
| Model \window setting | Full-range | Emphysema | Full-range | Emphysema |
| ResNet-34 | 0.75 [0.70, 0.80] | 0.79 [0.74, 0.84] | 0.78 [0.76, 0.80] | 0.82 [0.80, 0.84] |
| EfficientNetB2 | 0.80 [0.71, 0.90] | 0.89 [0.87, 0.91] | 0.75 [0.65, 0.85] | 0.79 [0.75, 0.83] |
| DenseNet-121 | 0.80 [0.76, 0.85] | 0.86 [0.82, 0.89] | 0.78 [0.73, 0.83] | 0.82 [0.80, 0.84] |

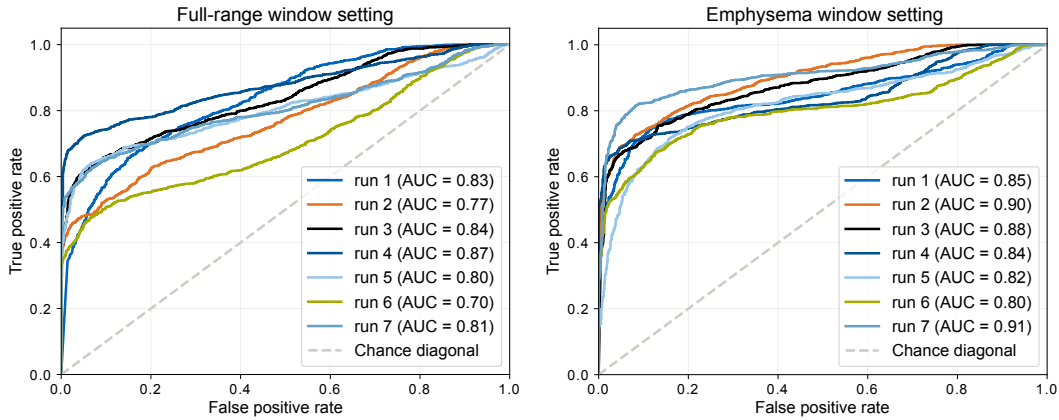

**Figure 4:** ROC plots and AUC values show inference on slice-level contrast-enhanced CT test data for each run of the plain DenseNet for input data clipped to full-range and emphysema window settings.

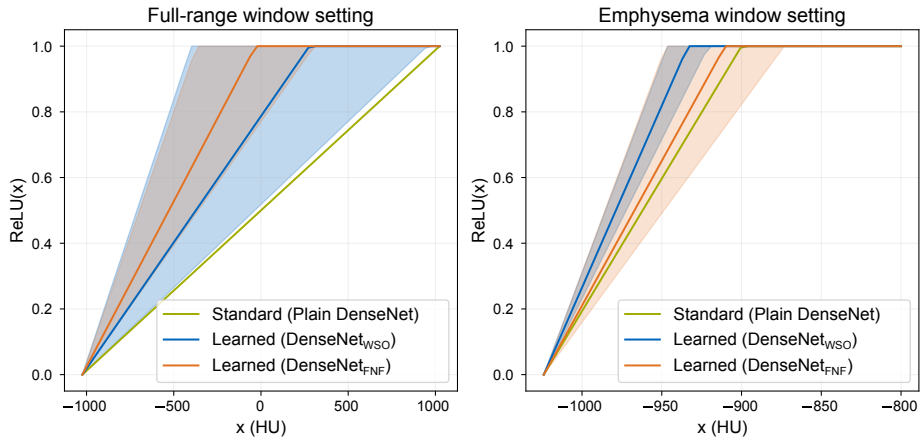

**Figure 5:** Learned window settings with 95% confidence intervals (CI). Standard full-range and emphysema window settings (green) are plotted against the mean learned window setting with 95% CI over seven runs for DenseNet$_{WSO}$ (blue) and DenseNet$_{FNF}$ (orange). Note that the standard window settings were used to preprocess the inputs for the plain DenseNet and to initialize the DenseNet$_{WSO}$ and DenseNet$_{FNF}$. The exact values for window settings are provided in Table 3.

**Table 3**

Mean Standard and Learned Window Setting by DenseNet Variants (n=7)

| Model \window setting | Full-range (Width, Level) HU | Emphysema (Width, Level) HU |
|---|---|---|
| Standard (Plain DenseNet) | (2048, 0) | (124, -962) |
| Learned (DenseNet$_{WSO}$) | (1301 [676, 1927], -373 [-686, -61]) | (90 [79, 102], -979 [-985, -973]) |
| Learned (DenseNet$_{FNF}$) | (993 [681, 1305], -528 [-684, -372]) | (114 [79, 148], -967 [-984, -950]) |

of the initialization window setting. However, when initialized to the full-range window, the DenseNet$_{FNF}$ arrived at the mean WW and WL parameters over seven runs with less deviation than when the model was initialized to the emphysema window. Overall, better mean AUC values are achieved when the learned window setting is closer to the standard emphysema window.

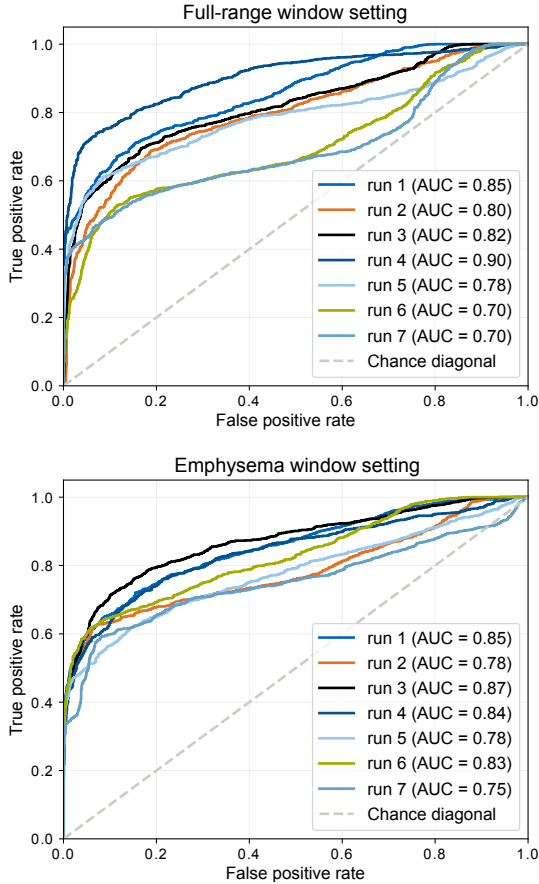

**Figure 6:** ROC plots and AUC values show inference on slice-level contrast-enhanced CT test data for DenseNet$_{WSO}$ models initialized to the full-range and the emphysema window settings for seven runs.

The ROC curves for the DenseNet$_{WSO}$ model and the DenseNet$_{FNF}$ model are depicted in Figure 6 and Figure 7 for CECT data, and in Figure S2 and Figure S3 for NCCT data, respectively. Generally, initialization to emphysema windowing results in more consistent AUC values over seven runs compared to the full-range window setting for both CECT and NCCT test data. Only for CECT data did the DenseNet$_{FNF}$ model generate more consistent AUC values over seven runs when initialized to the full-range window compared to the emphysema window setting. These results agree with the 95% CI values given in Table 3 and Figure 5 as the data was trained on CECT images. The highest AUC value achieved between the DenseNet$_{WSO}$ and the DenseNet$_{FNF}$ models was 0.91. This corresponded to the second run of the emphysema window setting initialization for the DenseNet$_{FNF}$ model given the CECT test data.

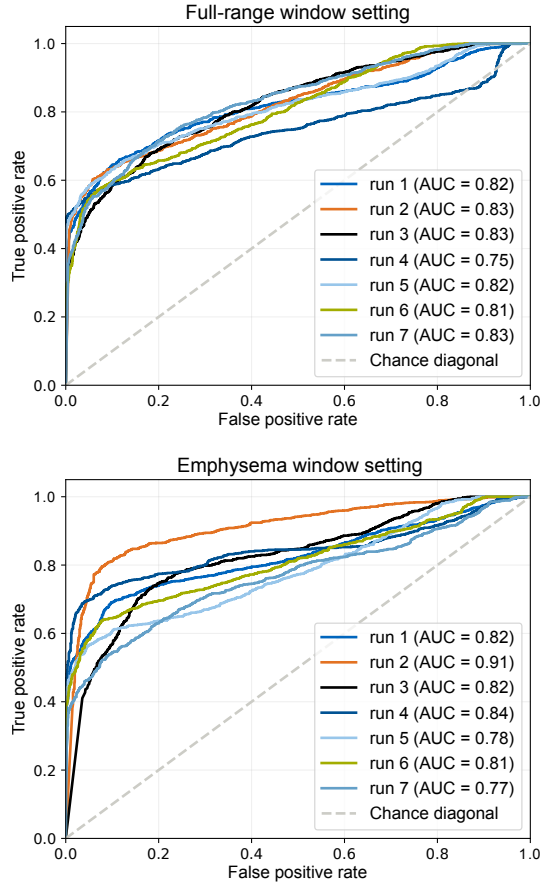

**Figure 7:** ROC plots and AUC values show inference on slice-level contrast-enhanced CT test data for DenseNet$_{FNF}$ models initialized to the full-range and the emphysema window settings for seven runs.

### 3.3. Optimal Window Setting

The mean AUC values for all model and window setting combinations for inference on the test sets are provided in Table 4. Taking the plain DenseNet model with full-range input images as the baseline for each test set (mean AUC$_{CECT}$ = 0.80 [0.76, 0.85], mean AUC$_{NCCT}$ = 0.78 [0.73, 0.83]), the plain DenseNet with input images initialized to the emphysema window setting showed best results (mean AUC$_{CECT}$ = 0.86 [0.82, 0.89], mean AUC$_{NCCT}$ = 0.82 [0.80, 0.84]) respectively. Implementing the WSO layer in DenseNet$_{WSO}$ and DenseNet$_{FNF}$ models did not drastically enhance the AUC compared to the results obtained with the plain DenseNet. Compared to the baseline, the DenseNet$_{FNF}$ model generated slightly better AUC values when initialized to either window setting. However, the most optimal window setting for the COPD detection task was the standard emphysema window setting of (124,-962) HU and not a window setting learned by either of the automated WSO models.

## 4. Discussion

We explored manually adjusted versus automated WSO for CT images with CNNs to optimize the binary detection of COPD. Manual preprocessing of CT images to emphysema window setting consistently improved binary detection of COPD with various CNNs. Specifically, DenseNet efficiently achieved a better mean $AUC_{CECT} = 0.86$ [0.82, 0.89] when provided with input data preprocessed to the emphysema window setting compared to data preprocessed to the full-range window setting. Furthermore, optimal window settings in the proximity of the emphysema window setting were learned by automating the window-setting optimization process through the addition of a customized layer to the DenseNet. Our findings demonstrate that diligent preprocessing based on existing radiological knowledge and selecting phenotypically representative ground truth labels positively impact the outcome of COPD detection with CNN models.

Three CNN architectures were examined based on their characteristics and existing results in literature: ResNets and DenseNets benefit from skip connections, improving the gradient stability and information flow throughout the network [30][32]. Furthermore, both models have shown promising results in COPD detection [14][15]. EfficientNets were also considered as they require less computing and promise fast training [31]. Although the results on the CECT test data show the best mean AUCs for EfficientNetB2, DenseNet-121 achieved comparable mean AUCs with a shorter mean training time, and demonstrated better and more robust outcome on the out-of-distribution NCCT test set. This behavior of the DenseNet-121 model is of particular interest when implemented in a clinical setting, where a timely initial fine-tuning of the model to new data distribution is advantageous. Additionally, the DenseNet-121 model has fewer trainable parameters compared to the EfficientNetB2, leading to more robustness during training and a lower likelihood of overfitting.

We showed that adjusting input images to different window settings directly impacts the binary detection of COPD; preprocessing CT images to the emphysema window consistently improves the performance of EfficientNetB2, ResNet-34, and DenseNet-121 models on the binary detection of COPD. Taking the AUC values for the plain DenseNet model with CECT full-range input data as the baseline, when the CECT input was preprocessed to the emphysema window, the mean AUC value increased from 0.80 to 0.86. The AUC values for the $DenseNet_{WSO}$ model suggest the shortcoming of the model in simultaneously detecting COPD and converging to optimal windowing parameters when initialized to the full-range window setting. Furthermore, the windowing parameters learned with this setup suffered from large deviations across the seven runs, as evidenced by the wide 95% CIs. To combat this, the WSO layer was trained with periodically frozen learnable parameters, as implemented in $DenseNet_{FNF}$. Improved mean AUC values were obtained with the $DenseNet_{FNF}$ models in comparison to the $DenseNet_{WSO}$ models, and narrower 95% CIs in learned window settings were observed when $DenseNet_{FNF}$ was initialized to the full-range window setting.

Through automatic WSO, only minimal improvement in AUC value was observed with the $DenseNet_{WSO}$ and the $DenseNet_{FNF}$ models initialized to emphysema window setting, compared to the baseline. The window settings learned by these two models were in the vicinity of the standard emphysema window at the lower ranges of the HU scale. However, neither $DenseNet_{WSO}$ nor $DenseNet_{FNF}$ outperformed the plain DenseNet model with images preprocessed to the standard emphysema window setting. A possible explanation is that although the single WSO layer converged to the optimal emphysema window setting, it was not sufficiently complex for optimal window setting selection. The main advantage of including a WSO layer is the additional information obtained regarding the window setting range appropriate for the task at hand. More specifically, when information regarding the optimal window setting for the detection of COPD is unknown, learning an optimal windowing by adding a WSO layer with the $DenseNet_{FNF}$ model results in higher AUC values in comparison to training a plain DenseNet model with full-range normalized images.

The standard emphysema windowing is tailored to present high contrast between healthy and emphysematous lung tissues. Therefore, as the ground truth labels for our dataset were graded based on the severity of emphysema, the results were in line with the hypothesis that images clipped directly to the standard emphysema windowing or automatically clipped with the WSO layer to a learned window-setting in the proximity of standard emphysema window, would improve detection of COPD with DenseNets. Optimizing for window setting to increase contrast in images was effective for the detection task because the Fleischner Score ground-truth labels were directly based on disease-relevant morphological changes in the lung. This is further implied by the results of comparing the CECT test set for Fleischner score and GOLD score ground truth labels. Utilizing the Fleischner Score as ground-truth labels also enabled us to achieve comparable results to related works in the literature, despite using a smaller dataset [12][13][14][15].

**Table 4**
Mean Area Under the Receiver Operating Characteristics Curve (AUC) for DenseNet Variant Comparison (n=7)

| Test Set | Contrast-enhanced CT | | Non-contrast CT | |
|---|---|---|---|---|
| Model \window setting | Full-range | Emphysema | Full-range | Emphysema |
| Plain DenseNet | 0.80 [0.76, 0.85] | **0.86 [0.82, 0.89]** | 0.78 [0.73, 0.83] | 0.82 [0.80, 0.84] |
| DenseNet$_{WSO}$ | 0.79 [0.73, 0.86] | 0.81 [0.78, 0.85] | 0.77 [0.72, 0.82] | 0.78 [0.76, 0.80] |
| DenseNet$_{FNF}$ | 0.81 [0.79, 0.83] | 0.82 [0.78, 0.86] | 0.79 [0.75, 0.83] | 0.80 [0.77, 0.83] |

This exploratory study has some limitations: All subjects were examined at the same hospital. More notably, the relatively small dataset instigated intra-image correlation for image-level evaluations. Furthermore, the progression of COPD and emphysema are commonly controlled with native NCCT data [8]. To alleviate some of these short-comings and further test the robustness of our proposed method on out-of-distribution data, we additionally tested all models on an NCCT test set consisting of 5,086 CT slices (2,543 COPD, 2,543 no COPD). The models expectedly showed higher AUC values for the CECT test set as they were also trained on CECT data. However, the results for the NCCT test data were comparable and demonstrated the same trend for the DenseNet variants. The following is a suggested explanation: We used a range of HU values for the emphysema window setting as opposed to a fixed thresholding approach, such as the defined standard emphysema threshold at -950 HU for NCCT [27]. Therefore, the impact of varied HU values induced by the presence of contrast material was minimal as the defined emphysema window covered a range of relevant HU values [35].

The extendibility of our findings to a larger, more diverse dataset should be further explored. In the context of COPD, future works could further apply the proposed method for the categorical classification of the disease based on the progression scale introduced by the Fleischner Society. Additionally, future models can integrate both GOLD and Fleischner scores as ground truth labels in the training process for simultaneous and accurate detection of COPD based on both clinically relevant definitions of the disease.

We showed that optimizing for a task-specific window-setting improved CNN outcome by enhancing disease-relevant information from the input data. Our findings can be extended to a range of computer vision tasks in medicine, focusing on X-ray and CT data. By incorporating disease-relevant window settings commonly used by radiologists into the deep learning pipeline, the performance of models can be improved.

## Data Availability

Due to patient privacy, the training and testing data is unavailable. However, all methods are described sufficiently to be replicated with other data.

## Supplementary Materials

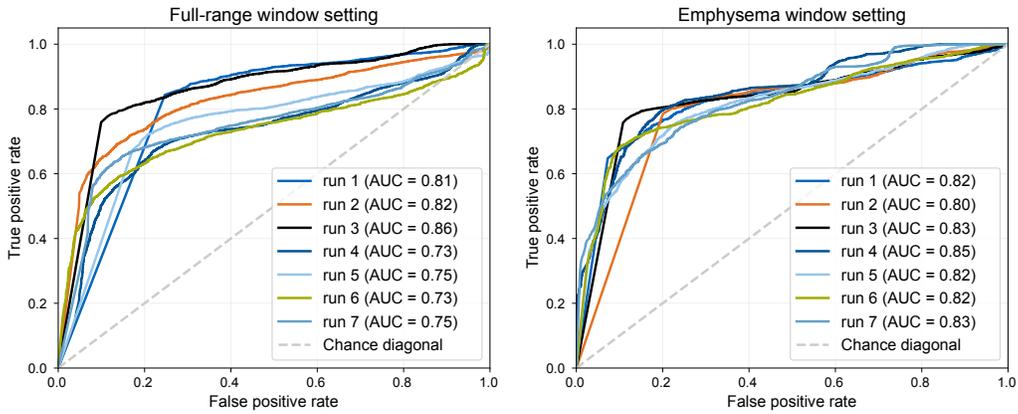

**Figure S1:** ROC plots and AUC values show inference on slice-level non-contrast CT test set for each run of the plain DenseNet for input data clipped to full-range and emphysema window settings.

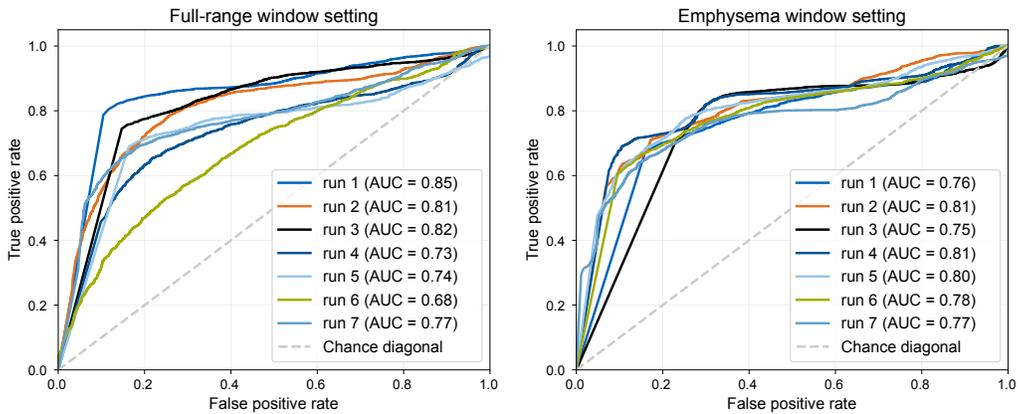

**Figure S2:** ROC plots and AUC values show inference on slice-level non-contrast CT test set for DenseNet$_{WSO}$ models initialized to the full-range and the emphysema window settings for seven runs.

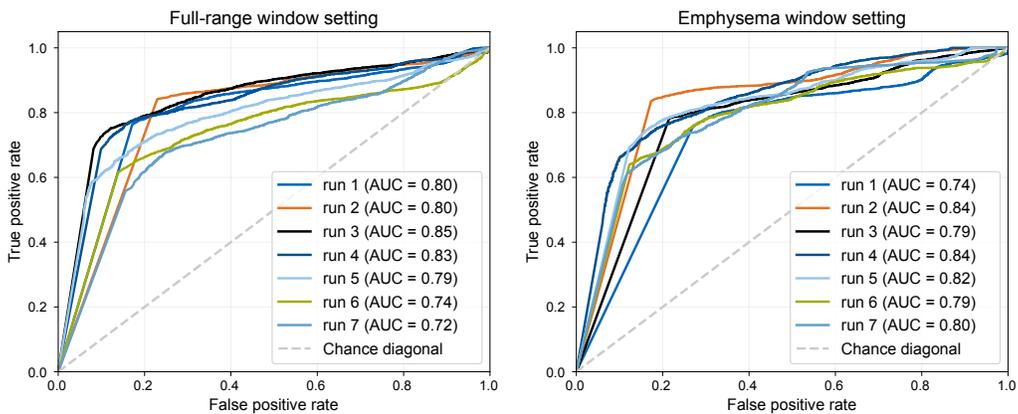

**Figure S3:** ROC plots and AUC values show inference on slice-level non-contrast CT test set for DenseNet$_{FNF}$ models initialized to the full-range and the emphysema window settings for seven runs.

**Table S1**

Mean Area Under the Receiver Operating Characteristics Curve (AUC) of the Contrast-enhanced Train and Validation CT Data for DenseNet Variants (n=7)

| Model \window setting | Train | | Validation | |
|---|---|---|---|---|
| | Full-range | Emphysema | Full-range | Emphysema |
| Plain DenseNet | 0.93 [0.86, 1.00] | 0.96 [0.92, 1.00] | 0.89 [0.84, 0.94] | 0.94 [0.90, 0.98] |
| DenseNet$_{WSO}$ | 0.94 [0.88, 1.00] | 0.90 [0.84, 0.96] | 0.86 [0.79, 0.93] | 0.87 [0.85, 0.89] |
| DenseNet$_{FNF}$ | 0.95 [0.87, 1.00] | 0.97 [0.95, 0.99] | 0.87 [0.82, 0.92] | 0.87 [0.82, 0.92] |

**Table S2**

Mean Area Under the Receiver Operating Characteristics Curve (AUC) of the Contrast-enhanced CT Test Data with GOLD Score Ground Truth for DenseNet Variants (n=7)

| Model \window setting | Full-range | Emphysema |
|---|---|---|
| Plain DenseNet | 0.88 [0.84, 0.92] | 0.87 [0.85, 0.89] |
| DenseNet$_{WSO}$ | 0.85 [0.80, 0.90] | 0.86 [0.83, 0.89] |
| DenseNet$_{FNF}$ | 0.88 [0.83, 0.93] | 0.85 [0.84, 0.86] |